\documentstyle[prd,tighten,aps]{revtex}

\def\R{{I\!\!\!\,R}}
\def\qPsi{|\Psi\rangle}
\def\beq{\begin{equation}}
\def\eeq{\end{equation}}
\def\bear{\begin{eqnarray}}
\def\ear{\end{eqnarray}}
\def\eql{&=&}
\def\bearr{\begin{eqnarray}&&}
\def\yyy{\\&&}
\def\nn{\nonumber\\}
\def\nnn{\nonumber\yyy}
\def\nnv{\vspace*{.1cm}\nn}
\def\d{\partial}
\def\e{{\rm e}}
\def\to{\rightarrow}
\def\sect{Sec.$\,$}
\def\eq{Eq.$\,$}
\def\eqs{Eqs.$\,$}
\def\nq{\hspace*{-.3cm}}
\def\cm{\hspace*{1.cm}}
\def\inch{\hspace*{1.in}}
\def\al{&}
\begin{document}

\twocolumn[\hsize\textwidth\columnwidth\hsize\csname
@twocolumnfalse\endcsname

\title{Geometric Interpretation of Thiemann's Generalized Wick
Transform}
\author{Guillermo A. Mena Marug\'{a}n}
\address{Instituto de Matem\'{a}ticas y F\'{\i}sica Fundamental,
C.S.I.C., Serrano 121, 28006 Madrid, Spain.}

\maketitle

\begin{abstract}
In the Ashtekar and geometrodynamic formulations of vacuum
general relativity, the Euclidean and Lorentzian sectors can be
related by means of the generalized Wick transform discovered by
Thiemann. For some vacuum gravitational systems in which there
exists an intrinsic time variable which is not invariant under
constant rescalings of the metric, we show that, after such a
choice of time gauge and with a certain identification of
parameters, the generalized Wick transform can be understood as
an analytic continuation in the explicit time dependence. This
result is rigorously proved for the Gowdy model with the
topology of a three-torus and for a whole class of cosmological
models that describe expanding universes. In these gravitational
systems, the analytic continuation that reproduces the
generalized Wick transform after gauge fixing turns out to map
the Euclidean line element to the Lorentzian one multiplied by
an imaginary factor; this transformation rule differs from that
expected for an inverse Wick rotation in a complex rescaling of
the four-metric. We then prove that this transformation rule for
the line element continues to be valid in the most general case
of vacuum gravity with no model reduction nor gauge fixing. In
this general case, it is further shown that the action of the
generalized Wick transform on any function of the gravitational
phase space variables, the shift vector, and the lapse function
can in fact be interpreted as the result of an inverse Wick
rotation and a constant, imaginary conformal transformation.
\end{abstract}

\vskip2pc]

\renewcommand{\thesection}{\Roman{section}}
\renewcommand{\thesubsection}{\Alph{subsection}}
\renewcommand{\theequation}{\arabic{section}.\arabic{equation}}
\renewcommand{\thefootnote}{\arabic{footnote}}
         %%%%%%%%%%%%%%%%%%%%%%%%%%%%%

\section {Introduction}

The application of the canonical quantization pro\-gram\-me to the
Ashtekar formulation of general relativity [1,2] is one of the most
promising approaches to constructing a quantum theory of gravity.
The Ashtekar variables are a densitized triad and a canonically
conjugate connection. In these terms, the gravitational
first-class constraints acquire a low-order polynomial form
suitable for quantization. Owing to the fact that the Ashtekar
connection is genuinely complex for Lorentzian metrics [1], one
seems nonetheless bound to impose complicated reality conditions
[1,3] in the quantization process in order to arrive at the
physical, Lorentzian theory. This problem does not appear
in the Euclidean sector of general relativity because the Ashtekar
variables can be defined as real ones in this case.

Considerable progress has recently been made on the canonical
quantization of gravity assuming Euclidean reality conditions
for the Ashtekar variables [4]. Remarkably, the results achieved
for Euclidean gravity turn out to be of significance for
Lorentzian general relativity because there exists a
well-established relation between both theories. This relation
has been found by Thiemann [5] who constructed a generalized
Wick transform (GWT) which maps the Euclidean constraint
functionals to the Lorentzian ones without preserving the
reality conditions. In particular, by means of this transform
one can obtain Lorentzian physical states starting with the
Euclidean quantum theory [5,6]. Procedures determining the
physical, Lorentzian inner product have also been suggested [5].
In this sense, the transform discovered by Thiemann allows one
to work only with real Ashtekar variables when quantizing
general relativity.

The role of the GWT is in fact analogous to that played by the Wick
rotation (WR) in other approaches to quantum gravity since they
both provide a relation between the Euclidean and Lorentzian
theories. In the path integral approach, e.g., it is assumed
that gravitational quantum states and correlation functions can
be defined by means of a sum over Euclidean histories, each of them
weighed by the exponential of minus the Euclidean action [7]. This
action, $I_{(E)}$, is obtained from the Lorentzian one, $S_{(L)}$,
by means of a WR. There are various ways of implementing this
rotation. For instance, one can carry out an analytic continuation
from the Lorentzian time coordinate $t^{(L)}$ to $-it^{(E)}$,
$t^{(E)}$ being the Euclidean time [7]. At least formally, this
procedure generalizes to gravity the WR of the time parameter
that is usually employed in ordinary quantum field theory. An
alternative prescription consists in performing an analytic
continuation of the lapse function $N$ to negative imaginary values
[8] (i.e., $N\to -iN$, with $N$ positive), so that the
diagonal time component of the metric is continued from the negative
to the positive real axis [9]. These two prescriptions can be
considered equivalent inasmuch as they both formally map the set of
all Lorentzian geometries to its Euclidean counterpart and lead to
the same expression for the gravitational Euclidean action $I_{(E)}$.

Explicitly, this action is given by
$I_{(E)}=-i R\circ S_{(L)}$ where $R$ denotes the WR.
For pure gravity, the standard Hilbert-Einstein action for
Lorentzian metrics is [7,10]
\bear
  S_{(L)} \eql \frac{1}{2}\int_{{\cal M}}d^4x\sqrt{-g}\,
        R\, -\int_{\d {\cal M}}d^3x\sqrt{h}\, k \nn
    \eql \frac{1}{2}\int_{{\cal M}}d^4x N \sqrt{h}\, (^{(3)}\!R
    +k_a^{\,b}
        k_b^{\,a}-k^2).
\ear
Here we have set $8\pi G=1$, $G$ being the gravitational
constant, $\d {\cal M}$ (supposed to consist of sections of
constant time) is the boundary of the four-manifold ${\cal M}$,
$k$ is the trace of the extrinsic curvature $k_a^{\,b}$, and
$g$, $R$, $h$, and $^{(3)}\!R$ denote the determinants and
curvature scalars of the four-metric and the induced
three-metric. A WR leads then to
\bear
I_{(E)} \eql \!\!-\,\frac{1}{2}\int_{{\cal M}}d^4x
        \sqrt{g}\, R\, -\int_{\d {\cal M}}d^3x \sqrt{h}\,k \nn
 \eql \frac{1}{2}\int_{{\cal M}}d^4x N \sqrt{h}\, (-^{(3)}\!R+
 k_a^{\,b}
    k_b^{\,a}\!-k^2).
\ear

Unlike the situation found in ordinary quantum field theory
in flat spacetime, where the Euclidean action is positive, the above
action is not even bounded from below [7,11]. Hence, Euclidean
path integrals for gravity are in principle ill-defined. It has
been argued that, in order to make these integrals converge
(at least in the one-loop approximation), the conformal factor must
be integrated over an appropriate complex contour other than the
real axis [11,12]. If it were possible to obtain meaningful path
integrals in this way, their analytic continuation
back to the Lorentzian section would supply physical results for the
Lorentzian theory [7].

The WR for gravity must not be considered to be a rigorously
defined transformation which sends each particular Lorentzian
metric to a real, Euclidean one. Indeed, it is known that the
analytic continuation of a Lorentzian metric does not generally
admit a section in the complexified spacetime on which the
metric has Euclidean signature and is real [13]. One must rather
understand the WR as a series of transformation rules for the
lapse function, the shift vector, and the gravitational phase
space variables (obtained, e.g., by any of the prescriptions
commented above) such that they formally map the Lorentzian
abstract line element to the Euclidean one and the Lorentzian
action for gravity to the Euclidean action multiplied by a
factor of $i$. It is in this sense that we will refer to the
gravitational WR from now on.

On the other hand, note that the WR is used to transform a sum
over Lorentzian configurations into a sum over Euclidean histories,
the latter expected to be more manageable. As we have said, it is
not assumed that each Lorentzian configuration has a real Euclidean
counterpart which can be reached by means of a WR; only the path
integrals over all configurations, either Lorentzian or Euclidean,
are supposed to be related [13]. A similar philosophy
underlies the introduction of the GWT. Instead of dealing with
Lorentzian gravity, whose reality conditions are very complicated,
one maps this theory to the Euclidean one. Furthermore, it is not
generally true that the GWT sends real solutions of the Euclidean
constraints to real Lorentzian solutions in the classical theory;
a correspondence between Euclidean and Lorentzian solutions is
established only quantum-mechanically [5,6].

However, there exists an important difference between the GWT
constructed by Thiemann and the WR. While the latter can be regarded
as an analytic continuation based on a complexification of the time
structure, the former is a map on functions on the phase space of
general relativity whose definition does not seem to rest on the
availability of a time coordinate and which results in complexifying
the Ashtekar variables [5]. Given the analogies between the GWT and
the WR, one might nonetheless expect that there exists a close
relation between the actions of these two transformations. In this
work, we find this relation and show that it allows one to reach a
geometric interpretation of the GWT. In fact, there has been some
confusion in the literature about the existence of such a relation.
When Thiemann introduced his transform, he claimed that its action
was a phase space WR [5], even though the GWT preserves Poisson
brackets, while the WR does not. In addition, it was suggested in
Ref.\,[6] that a simple spacetime interpretation of the GWT in terms
of a WR could not exist; in particular, it was argued that the lapse
function and the shift vector transform under the GWT in a way that
is different from what an interpretation of this kind would suggest.
One of the purposes of our analysis is to clarify this
point. We will see that the effect of the GWT in vacuum gravity can
really be interpreted as the result of an inverse WR composed
with a complex conformal transformation.

With the aim at gaining insight into the relation between the
GWT and the WR, we will first study some particular
gravitational systems. In doing this, we also want to discuss
another issue, namely, whether, under appropriate circumstances,
the GWT can further admit the spacetime interpretation of an
analytic continuation in the explicit time dependence that could
be related to the usual prescription for carrying out the WR. In
fact, one would expect this to be the case in the following
situation.

Given a vacuum gravitational
system, one can always find a canonical set of variables such that
the configuration variables are homogeneous functions of (only)
the triad. The degree of homogeneity can always be chosen to be
zero, except for one of the configuration variables, $x$. Taking
into account that the GWT preserves Poisson brackets and rescales
the triad by a constant complex factor [5], it is then possible to
prove that there exists a choice of canonical momenta such that the
only phase space variables affected by the GWT are the canonical
pair $(x,p_x)$, and that $x$ gets multiplied by a complex constant
under the GWT, while $xp_x$ remains invariant. It hence turns out
that the effect of the GWT on the three-metric (and, subsequently,
on the line element) amounts to a rescaling of $x$. Suppose then
that, for our particular gravitational system, $x$ is a positive
variable in the sector of non-degenerate metrics and that
$x=\tau\in \R^+$ is an acceptable choice of intrinsic time
($\tau$ being the time coordinate). This would occur, e.g.,
when $x$ describes a volume, area, or length element expanding
with the evolution. When the above conditions are satisfied, one
would expect that, after gauge fixing, the action of the GWT on the
line element could be viewed as a constant complex rescaling in the
explicit $\tau$-de\-pendence. This rescaling can be interpreted as an
analytic continuation in the complex $\tau$-plane. If one wants
to deal only with variables defined over the whole real axis, one
can replace $(x,p_x)$ with $(z=\ln{x},p_z=xp_x)$, and fix
$z=t=\ln{\tau}\in \R$. In terms of the time coordinate $t$,
the action of the GWT would be interpretable as a constant
complex translation in the explicit $t$-de\-pendence.

The situation described above occurs in fact in the two kinds of
vacuum gravitational systems that we are going to analyze.
We will first consider the family of Gowdy cosmologies with the
spatial topology of a three-torus [14], a model with two commuting
spacelike Killing vector fields. In this model, the surfaces that
contain the orbits of the Killing vector fields are in continuous
expansion. Hence, their area provides a natural choice for the
intrinsic time variable $x$. The other case that we will study is
that of expanding universes. By these we understand spacetimes such
that the trace of the extrinsic curvature is nowhere vanishing. The
sections of constant time expand then with the evolution, and we can
use their volume to define the intrinsic time variable $x$.

The rest of the paper is organized as follows. In \sect 2 we
summarize some basic results for Euclidean and Lorentzian gravity,
both in the triad and Ashtekar formulations, and review the
definition of the GWT. \sect  3 is devoted to the Gowdy cosmologies.
The case of expanding universes is studied in \sect  4. For these two
types of gravitational systems we prove that, after gauge fixing
and with a certain identification of parameters, the GWT can indeed
be interpreted as an analytic continuation performed exclusively in
the explicit time dependence and which resembles an inverse WR. The
action of the GWT and its geometric interpretation in the case of
general relativity without any gauge fixing nor model reduction
are discussed in \sect  5. We show that the GWT has the same
effect on the abstract line element and the gravitational action
as the composite of an inverse WR and a rescaling of the metric by
an imaginary factor. Moreover, we show that the action of the GWT
on the phase space variables, the shift vector, and the densitized
lapse function can be regarded as the result of an inverse WR and a
complex conformal transformation provided that this WR is implemented
by rotating the lapse function. We present our results and conclude
in \sect  6. Finally, some calculations that are useful for studying
expanding universes are worked out in the Appendix.

\section{The generalized Wick transform}
\setcounter{equation}{0}

In the triad formulation of general relativity, one can take as
elementary phase space variables the densitized triad
$\tilde{E}_i^a$ and the
contraction of the extrinsic curvature with the triad, $K_a^i$.
These variables are defined on a three-manifold $\Sigma$ and can
be restricted to be real. Lower-case Latin indices from the
beginning and the middle of the alphabet stand for spatial and SU(2)
indices, repectively. The latter are raised and lowered with the
identity metric. The fundamental Poisson brackets are
\beq
    \{\tilde{E}_i^a(x),K_b^j(y)\}=\delta_b^a\delta_i^j
        \delta^{(3)}(x-y)
\eeq
where $\delta_b^a$ and $\delta_i^j$ are Kronecker deltas, $x$ and
$y$ two points of $\Sigma$, and $\delta^{(3)}$ the delta function on
$\Sigma$.

Assuming that the induced metric $h_{ab}$ is not degenerate,
the variables $\tilde{E}_i^a$ and $K_a^i$ admit the following
expression in terms of the triad $e^a_i$ and the extrinsic
curvature:
\beq  \tilde{E}_i^a=\sqrt{h}\,e^a_i,\;\;\;\;\;
K_a^i=k_{ab}e^b_i.\eeq
Here $h$ is the determinant of $h_{ab}$, with $h^{ab}=e^a_ie^{bi}$.

The Gauss and vector gravitational constraints [6] read,
respectively,
\beq
    {\cal G}_i\equiv \epsilon_{ij}^{\;\;\;\;k}K_a^j
    \tilde{E}_k^a=0,\;\;\;\;\;{\cal V}_a\equiv 2\tilde{E}_i^b D_{[a}
    K_{b]}^i=0,
\eeq
where $\epsilon_{ijk}$ is the antisymmetric symbol,
the brackets denote antisymmetrization, and $D_a$ is the derivative
operator defined by $D_a\tilde{E}_i^b=0$. In particular, we
have [15]
\beq
    D_aK_b^i=\d _aK_b^i-\Gamma^c_{ab}K_c^i
    +\epsilon^i_{\;\;jk}\Gamma_a^jK_b^k.
\eeq
In this formula, $\Gamma^c_{ab}$ are the Christoffel symbols [16],
and $\Gamma_a^i$ is the spin connection compatible with the triad
[15], namely,
\beq
    \Gamma_a^i=-\frac{1}{2}\epsilon^{ijk}
    E_{_{_{\!\!\!\!\!\!\sim}}\;jb} (\d _a\tilde{E}^b_k+
    \Gamma^b_{\;ca}\tilde{E}^c_k),
\eeq
$E^i_{_{_{\!\!\!\!\!\!\sim}}\;a}$ being the inverse of the
densitized triad.

The only remaining first-class constraint is the scalar one,
which is different for Lorentzian and Euclidean metrics [6].
We can nevertheless consider both cases simultaneously by
introducing a parameter $\epsilon$ such that
\beq
\epsilon=\left\{ \begin{array}{lc} -i & {\rm for\; Lorentzian
            \;metrics} \\
\;\,1 & {\rm for\; Euclidean\; metrics}\end{array} \right.
\eeq
The scalar constraint can then be written in the form
\beq  {\cal S}\equiv 2\tilde{E}_{\,i}^{\,[a}
\tilde{E}_j^{b]}K_a^iK_b^j-\frac{h}{\epsilon^2} \;
^{(3)}\!R=0.
\eeq

The Ashtekar variables can be obtained from $\tilde{E}_i^a$ and
$K_a^i$ by means of a canonical transformation which results
in replacing the momenta $K_a^i$ with the connection [1,2]
\beq
    A_a^i=\Gamma_a^i+\epsilon\,K_a^i.
\eeq
Note that the Ashtekar connection $A_a^i$ is real for Euclidean
gravity but complex in the Lorentzian case. The Poisson-bracket
structure is given by
\beq
    \biggl\{\frac{1}{\epsilon}\tilde{E}_i^a(x),A_b^j(y)\biggr\}=
    \delta^a_b\delta^j_i\delta^{(3)}(x-y),
\eeq
and the gravitational constraints can be expressed as [15]
\bearr
    {\cal G}_i=\frac{1}{\epsilon}{\cal D}_a
    \tilde{E}_i^a=\frac{1}{\epsilon}(\d _a\tilde{E}_i^a+
    \epsilon_{ij}^{\;\;\;\;k}A_a^j\tilde{E}_k^a)=0,\yyy
  {\cal V}_a\approx
    \frac{1}{\epsilon}\tilde{E}_i^bF^i_{ab}=0\yyy
  {\cal S}\approx\frac{1}{\epsilon^2}
        \epsilon^{ij}_{\;\;\;\;k}
        \tilde{E}_i^a\tilde{E}_j^bF^k_{ab}=0,
\ear
where the symbol $\approx$ denotes equality modulo the Gauss
constraint, ${\cal D}_a$ is the gauge-covariant derivative defined
by the Ashtekar connection, and $F^i_{ab}$ is its curvature:
\beq
    F^i_{ab}=\d _aA_b^i-\d _bA_a^i+
        \epsilon^i_{\;\;jk}A_a^jA_b^k.
\eeq

In order to introduce the GWT, let us first suppose that
$C(t,\rho)$ is a given function on the gravitational phase
space which may explicitly depend on a certain parameter $t$,
$\rho$ denoting a complete set of phase space variables.
Using $C$ as an infinitesimal generator, we can construct the
following family of maps on functions $f(\rho)$ on phase space:
\beq
    W(t, t_0)\circ f=
    \sum_{n=0}^{\infty}\frac{1}{n!}\,{\cal P}\left[\{f,
    \int^t_{t_0}C\}_{(n)}\right].
\eeq
Here, $\{\,,\,\}_{(n)}$ is the multiple Poisson bracket [5], and
the symbols ${\cal P}[\int^t_{t_0}]$ denote path-ordered
integration over the explicit dependence on $t$, the contour of
integration being the segment that joins $t_0$ with $t$ (in the
complex $t$-plane) and $t_0$ being a fixed initial value of our
parameter. Explicitly we can write the $n$-th term of the above
expression as [17]
\[ \nq\,
    \Bigl\{\Bigl\{...\Bigl\{f(\rho),\int_{t_0}^t dt_n...
        \int_{t_0}^{t_2}dt_1 C(t_1,
            \rho)\Bigr\}...\Bigr\},C(t_n,\rho)\Bigr\}.
\]
Assuming that the series in Eq. (2.14) converges, each map
$W(t,t_0)$ is an automorphism on the algebra of functions on
phase space that preserves the Poisson-bracket structure. However,
$W(t,t_0)$ generally modifies the reality conditions if the contour
of integration for $t$ or the function $C$ along this contour are
not real [5,6].

In the case that $C$ is explicitly independent of $t$, Eq. (2.14)
reduces to
\beq
    W(t-t_0)\circ f=
        \sum_{n=0}^{\infty} \frac{(t-t_0)^n}{n!}\,
    \{f,C\}_{(n)}.
\eeq
The GWT is defined as the map $W\equiv W(t-t_0=i\pi/2)$ obtained
with the infinitesimal generator
\beq
    C=\int_{\Sigma} d^3x\, K_a^i \tilde{E}_i^a.
\eeq
From \eqs (2.1) and (2.15) it is straightforward to check that
\beq
    W\circ \tilde{E}_i^a= i \tilde{E}_i^a,\cm
    W\circ K_a^i=-i K_a^i.
\eeq
Since the spin connection $\Gamma_a^i$ is a function of only the
densitized triad that is invariant under constant rescalings of
$\tilde{E}_i^a$, we then obtain that the GWT maps the Euclidean
Ashtekar connection to the Lorentzian one. Moreover, one can now
readily see that the GWT maps the Euclidean constraints
(2.10)-(2.12) (with $\epsilon=1$) to those of Lorentzian gravity
(i.e., those with $\epsilon=-i$):
\bearr
    W\circ{\cal G}_i^{\,(E)}={\cal G}_i^{\,(L)},\cm
    W\circ{\cal V}_a^{\,(E)}={\cal V}_a^{\,(L)},   \nnn
        W\circ{\cal S}^{\,(E)}={\cal S}^{\,(L)}.
\ear
The superscripts $(E)$ and $(L)$ are used in this formula to refer
to the Euclidean and the Lorentzian theory, respectively.
The above conclusion about the transformation of the constraints
can also be reached from the alternative expressions (2.3) and (2.7)
by using the fact that the first equation in (2.17) implies the
constant conformal transformation $W\circ h_{ab}=i h_{ab}$ and by
realizing that
\bearr
    ^{(3)}\!R(i h_{ab})=-i\;^{(3)}\!R(h_{ab}), \nnn       %% 2.19
    \det (i h_{ab})=i^3\,h.
\ear

To pass to quantum theory, let us assume that the infinitesimal
generator $C$ has a well-defined quantum analogue $\hat{C}$, and
(setting $\hbar=1$) introduce the operator
\beq
    \hat{W}= {\cal P} \left[\exp{\left(i\int_{t_0}
    ^{t_0+i\frac{\pi}{2}}\hat{C}\right)}\right]=
    \exp{\left(-\frac{\pi}{2}\hat{C}\right)}.
\eeq
The classical map $W$ has then the quantum mechanical counterpart [5]
\beq
    \hat{W}\,\hat{f}\,\hat{W}^{-1}=\sum_{n=0}^{\infty}
    \left(\frac{\pi}{2}\right)^n \frac{\,[\hat{f},\hat{C}]_{(n)}}{n!}
\eeq
where $\hat{f}$ is a generic operator. From \eq (2.18) it
follows that $\hat{W}$ transforms Euclidean quantum constraints
into Lorentzian ones (with an appropriate choice of factor
ordering). This implies that a quantum state $\qPsi $ is
annihilated by the constraints of Euclidean gravity if and only
if $\hat{W}\qPsi$ is (at least formally) a solution to the
Lorentzian constraints [5,6]. In this way the physical states of
Lorentzian general relativity can actually be obtained from the
Euclidean quantum theory.

\section{The Gowdy model}
\setcounter{equation}{0}

To gain insight into the kind of relation that can exist between
the GWT and the WR, we will first consider the action of the GWT
in a particular gravitational system, namely, the family of
Gowdy cosmologies whose sections of constant time have the
topology of a three-torus. This gravitational model has been
studied in Ref.\,[18], both clasically and quantum-mechanically,
although the analysis carried out there was restricted to the
sector of non-degenerate Lorentzian metrics. We first review
that analysis and extend it to the case of Euclidean metrics.
The GWT for the model is studied at the end of this section. We
will show that there exists a choice of intrinsic time such that
the GWT can be interpreted as an analytic continuation made in
the explicit time dependence. This continuation turns out to map
the Euclidean line element of the model to its Lorentzian
counterpart multiplied by a factor of $i$. We will also comment
on the validity of these results for other choices of time. In
particular, we will argue that the conclusion that the GWT maps
the Euclidean line element to the Lorentzian one rescaled by a
complex conformal factor is gauge-independent.

\subsection{Lorentzian and Euclidean models}

The Gowdy universes are vacuum spacetimes which possess two
commuting spacelike Killing vector fields and whose sections of
constant time are compact [14]. When these sections have the
topology of a three-torus, one can choose a global
set of coordinates $(t,\theta,\omega,\nu)$ such that
$\d _{\omega}$ and $\d _{\nu}$ are the two Killing
fields and $2\pi\omega,\, 2\pi\nu,\, \theta\in S^1$ ($S^1$ being
the unit circle).

It is possible to remove almost all of the non-physical degrees of
freedom of the model by a gauge-fixing procedure. One starts by
imposing the gauge-fixing conditions [18,19]
\beq
    \tilde{E}_M^{\theta}=\tilde{E}^{\alpha}_3=
    \arctan{\left(\frac{\tilde{E}^{\nu}_1}
    {\tilde{E}_2^{\nu}}\right)}=0.
\eeq
Capital Latin letters from the middle of the alphabet denote
SU(2) indices equal to 1 or 2, whereas Greek letters refer to
the spatial variables $\omega$ and $\nu$. The Gauss and vector
constraints ${\cal G}_M$ and ${\cal V}_{\alpha}$ are then solved
by
\beq
    A_{\theta}^M=A_{\alpha}^3=0,
\eeq
while the remaining Gauss constraint ${\cal G}_3$ implies
\beq
    \epsilon_M^{\;\;N3}A_{\alpha}^M\tilde{E}_N^{\alpha}
    +\d _{\theta}\tilde{E}_3^{\theta}=0.
\eeq
Using \eqs (2.2), (2.5), and (2.8), the conditions (3.1)--(3.3)
can be seen equivalent to the demand that the extrinsic
curvature be symmetric, and that all the non-diagonal elements
of the induced metric and the extrinsic curvature vanish, except
$h_{\omega\nu}$ and $k_{\omega\nu}$. After the above partial
gauge fixing, one can set the $\omega$ and $\nu$ components of
the shift vector equal to zero in the four-dimensional metric
[18].

Let us next rename $E=\tilde{E}^{\theta}_3$ and define, both for
the Lorentzian and Euclidean theories,
\bearr
    {\cal K}=\frac{1}{\epsilon}(A^3_{\theta}-\Gamma^3
    _{\theta}),\;\;\;\;\;K^M_{\alpha}=\frac{1}{\epsilon}
    (A_{\alpha}^M-\Gamma^M_{\alpha}),
\yyy
    \bar{K}_{\alpha}^{\,\beta}=K_{\alpha}^M
    \tilde{E}^{\beta}_M,\;\;\;\;\;q^{\alpha\beta}=
    \tilde{E}^{\alpha}_M
    \tilde{E}^{\beta M},
\ear
with $\epsilon$ being the parameter given by \eq (2.6). In the
sector of non-degenerate induced metrics, a canonical set of
real phase space variables for the reduced Gowdy model obtained
with our previous gauge fixing is [18]
\bear
u=2\ln{\left(\frac{\sqrt{{\rm det}
    (q^{\alpha\beta})}}{E q^{\nu\nu}}\right)},&\;\;\;
            & v=\frac{q^{\omega\nu}}{q^{\nu\nu}},\\
    z=\ln{E},\hspace*{2.25cm}& & w=\frac{1}{2}\ln{q^{\nu\nu}},\\
p_u=\frac{1}{2}(\bar{K}_{\omega}^{\,\omega}-\bar{K}_{\omega}
        ^{\,\nu}v),\hspace*{.4cm}& &p_v=\bar{K}_{\omega}^{\nu},\\
H_r=-{\cal K}E-2p_u,\hspace*{.8cm}& &
        \bar{K}=\bar{K}_{\alpha}^{\,\alpha}.
\ear
These variables depend only on $\theta\in S^1$ (and on the time
coordinate $t$). Their non-vanishing Poisson brackets are
\bear
    \{u(\theta),p_u(\theta^{\prime})\}\!
 \eql \!
 \{v(\theta),p_v(\theta^{\prime})\}=\{H_r(\theta),
z(\theta^{\prime})\}\nnv
 \eql
 \!\{w(\theta),\bar{K}(\theta' )\}
        =\delta(\theta-\theta' )
\ear
where $\delta(\theta)$ is the Dirac delta on $S^1$. This reduced
model still possesses two first-class constraints, namely,
the scalar one and the vector constraint ${\cal V}_{\theta}$.

It is worth commenting that, from our definitions, the phase
space variables $u$, $v$, $w$, and $z$ are the same for the
Lorentzian and Euclidean sectors, whereas their canonically
conjugate momenta ($p_u$, $p_v$, $\bar{K}$, and $-H_r$) differ,
in the Euclidean theory, from their Lorentzian counterparts by a
factor of $-i$. Except for what concerns the Poisson-bracket
structure and the reality conditions, this relation between the
two sets of phase space variables allows one to translate the
results for the Lorentzian model to the Euclidean case. This
prescription can be employed in what follows to check the
consistency of our analysis.

On the other hand, using \eq (2.2) and recalling that
$h_{\theta\alpha}=k_{\theta\alpha}=0$, it is not difficult to show
that
\beq
    \bar{K}=h\,k_{\alpha\beta}h^{\alpha\beta},\cm
        E=\sqrt{ {\rm det}(h_{\alpha\beta})}.
\eeq
Thus the variables $\bar{K}$ and $E$ are
proportional to and have the same sign as the trace of the
extrinsic curvature and the area element of the surfaces of constant
$\theta$-coordinate and time, respectively. In addition, it is known
that
\beq
    \bar{K}_0=\frac{1}{\sqrt{2\pi}}\oint \bar{K}
\eeq
(where the symbol $\oint$ denotes integration over $\theta\in S^1$)
is a constant of motion of the system, that is, its value is
preserved both by the dynamical evolution and by all gauge
transformations [20]. Thanks to this fact, one can consistently
restrict all considerations to the family of geometries with
non-vanishing $\bar{K}_0$ [18]. Moreover, taking into account the
invariance of the four-geometries under time reversal, one can
further restrict $\bar{K}_0$ to be positive without loss of
generality [18]. One can then use the $\theta$-diffeomorphism
gauge freedom to make $\bar{K}$ coincide with its average on
$S^1$, namely,
\beq
    \bar{K}=\frac{\bar{K}_0}{\sqrt{2\pi}}.
\eeq
The corresponding classical solutions describe universes with
sections of constant time and $\theta$-coordinate that expand
forever from an initial singularity. Actually, these solutions
are those analyzed in the Lorentz\-ian theory by Gowdy for the
topology of a three-torus [14, 18]. The area of the surfaces
with constant $t$ and $\theta$ provides hence a natural
candidate for defining an intrinsic time. We can then set, e.g.,
\beq
    z=t.
\eeq
This choice of intrinsic time is in fact equivalent to that made by
Gowdy [14], which can be expressed as $\tau=E$. From \eqs (3.7) and
(3.14) we then get $\tau=\e^t>0$. On the other hand, it is worth
noticing that, since $E$ scales as the densitized triad and is
positive for non-degenerate metrics, our choice of time satisfies
the conditions that, according to our discussion in the
Introduction, should guarantee that the GWT can be interpreted
after gauge fixing as an analytic continuation in the explicit time
dependence. We will see later in this section that such a spacetime
interpretation is actually feasible.

\eqs (3.13) and (3.14) can be imposed as gauge-fixing conditions to
eliminate almost all of the non-physical degrees of freedom of our
reduced model [18]. With this gauge fixing, the scalar constraint
can be solved to obtain an expression of $H_r$ in terms of
the time coordinate and the variables $\bar{K}_0$, $u$, $p_u$, $v$,
and $p_v$. Besides, the inhomogeneous part of the vector constraint
${\cal V}_{\theta}$ implies
\beq
    w=\frac{w_0}{\sqrt{2\pi}}-\,\sum_{\!\!n=-\infty,\neq
    0}^{\infty} \frac{\Pi_n}{in\bar{K}_0}\,\e^{in\theta},
\eeq
whereas its homogeneous part, which reads $\Pi_0=0$, remains as
the only constraint of the system [18]. Here $\Pi_n=\oint \Pi
\e^{-in\theta}/\sqrt{2\pi}$ are the Fourier coefficients of
\beq
    \Pi=\d _{\theta}u\;p_u+\d _{\theta}v \;p_v.
\eeq

In \eq (3.15) the zero-mode $w_0$ is left undetermined. This mode
is a homogeneous degree of freedom of the reduced model attained
with our gauge fixing. Its canonically conjugate momentum is
$\bar{K}_0$. Since $\bar{K}_0\in \R^+$, we can replace
the canonical pair $(w_0,\bar{K}_0)$ by
\beq
    b_0=\bar{K}_0w_0,\cm   c_0=\ln{\bar{K}_0}.
\eeq
The domain of definition of these variables is the entire real
axis, and they satisfy $\{b_0,c_0\}=1$. Finally, the conditions
(3.13) and (3.14) are compatible with the dynamical evolution
provided that the densitized lapse function
$N_{_{_{\!\!\!\!\!\!\sim}}\;}$ is given by
$\sqrt{2\pi}/\bar{K}_0$, and the $\theta$-component of the shift
vector $N^{\theta}$ depends only on time [18]. A suitable
redefinition of the $\theta$-coordinate sets then $N^{\theta}$
equal to zero [18].

The reduced model that results from our gauge fixing admits, as a
canonical set of real variables, the set formed by $b_0$, $c_0$ and
the fields on the unit circle $u$, $p_u$, $v$, and $p_v$. There
exists a homogeneous constraint: $\Pi_0=0$. The time evolution
is generated in the model by the reduced Hamiltonian density $H_r$
(i.e., minus the momentum of the time variable) obtained by solving
the scalar constraint. On the other hand, to check that the
Hamiltonian density $H_r=-{\cal K}E-2p_u$ coincides in the
Lorentzian case with that found in Ref.\,[18], it suffices to use the
definition of ${\cal K}$ and the fact that, once the conditions (3.1)
are imposed,
\beq
    \Gamma_{\theta}^3 E=-\frac{1}{2}\e^{-u/2} \d _{\theta}v.
\eeq

The canonical quantization of this reduced model has been discussed
in Ref.\,[18] by choosing as elementary variables $b_0$, $c_0$ and
the Fourier coefficients $(u_n,p_u^n,v_n,p_v^n)$ of the fields
$(u,p_u,v,p_v)$. The quantum states were represented by analytic
functionals $\Psi$ of the variables $(c_0,u_n,v_n)$ $[n=0, \pm 1,
...]$. These states can also depend on time. For the analysis of
the GWT, however, it will prove convenient to
make the following change of elementary variables:
\beq  u' _0=u_0+2\sqrt{2\pi}\,t,\;\;\;\;\;
b_0' =b_0-\sqrt{2\pi}\e^{c_0}t,\eeq which is just a canonical
transformation generated by the function
\beq  F=p_u^0(2\sqrt{2\pi}\,t-u_0' )-c_0
b_0' -\sqrt{2\pi}\e^{c_0}t.\eeq The momenta of $u_0' $ and $b_0'
$ are then $p_u^0$ and $c_0$, respectively. Notice that the
constraint $\Pi_0=0$ is invariant under the transformation
(3.19) because $\Pi$, given by \eq (3.16), does not depend on
$u_0$ nor $b_0$. With this change of variables, the line element
of our gauge-fixed model can be written in the form [18]
\bearr
    ds^2=\e^t \e^{2w' +u'/2}
    \left[\frac{2\pi}{\epsilon^{2}}\, \e^{2t} \e^{-2c_0} dt^2
        +d\theta^2\right]      \nnn
    +\e^t \e^{-u'/2}\left[d\omega^2-2vd\omega d\nu+
            (\e^{u' }+v^2)d\nu^2\right]
\ear
where
\bear
  u' \eql \frac{u_0' }{\sqrt{2\pi}}
    +\sum_{n\neq 0}\frac{u_n}{\sqrt{2\pi}}\e^{in\theta}=u+2t, \\
  w' \eql
    \e^{-c_0}\left(\frac{b_0' }
    {\sqrt{2\pi}}-\sum_{n\neq 0}\frac{\Pi_n}{in}\e^{in\theta}\right)
        =w-t,
\ear
the last equality coming from \eqs (3.15) and (3.17).

Since the function (3.20) depends on time, the dynamical evolution
of our new set of variables is generated by the reduced Hamiltonian
\bear
    H^T&\equiv& \oint H_r' =\oint H_r
                                +\d _t F\nn
    \eql \oint\left(H_r+2 p_u-\frac{\e^{c_0}}{\sqrt{2\pi}}\right)
\ear
where we have used the definition of the Fourier coefficient
$p_u^0$. Therefore we get that, with our gauge fixing,
\beq
    H_r' =-\left({\cal K}E+\frac{\bar{K}_0}
        {\sqrt{2\pi}}\right).
\eeq
On the other hand, substituting the solution obtained for $H_r$
from the scalar constraint [18] into \eq (3.24), we arrive at
the explicit expression
\bearr
 H_r' =-\sqrt{2\pi}\e^{-c_0}\left[4\left(p_u-\frac{\e^{c_0}}
            {4\sqrt{2\pi}}\right)^2+\e^{u' }p_v^2\right]\nnn
 -\frac{3\e^{c_0}}{4\sqrt{2\pi}}+\frac{\sqrt{2\pi}
    \e^{-c_0}}{16\epsilon^2}\e^{2t}\!\left[(\d _{\theta}
        u' )^2\!+4\e^{-u' }(\d _{\theta}v)^2\right]. \nnn
\ear

To conclude this subsection, let us comment that, starting from
the quantum representation discussed for our model in Ref.\,[18], it
is possible to construct a representation in which the quantum
states are analytic functionals of the set of variables
$(c_0,u_0' ,u_m, v_n)$ where $m=\pm 1, \pm 2, ...$ and
$n=0,\pm 1,...$. For each quantum state $\Psi$ in the former
representation, define
\bearr
    \Phi(c_0,u_0' ,u_m,v_n,t)\!=\!
    \Psi(c_0,u_0\!=\!u_0' \!-\!2\sqrt{2\pi}t,u_m,v_n,t)\nnn
         \cm   \times\;\exp{(i\e^{c_0}\sqrt{2\pi}t)}.
\ear
Under this correspondence, the states
$\hat{u}' _0\Phi\equiv u' _0\Phi$ and
$\hat{b}' _0\Phi\equiv i\d _{c_0}\Phi$ are mapped in the
original representation to
\bear
    u' _0\Phi \al \to \al (u_0+2\sqrt{2\pi}\,t) \Psi,\nn
            i\d _{c_0}\Phi
    \al \to \al (i\d _{c_0}- \sqrt{2\pi}\e^{c_0}t)\Psi.
\ear
These transformations are the quantum counterpart of the
relations (3.19). Moreover, since the only constraint of the
system has been represented in Ref.\,[18] by an operator
$\hat{\Pi}_0$ which involves neither multiplication by $u_0$ nor
differentiation with respect to $c_0$, it follows that $\Phi$ is
annihilated by $\hat{\Pi}_0$ if the same happens to $\Psi$. In
addition, it is straightforward to check that $\Psi$ is a
solution to the Schr\"{o}dinger equation $i\d _t\Psi=\hat{H}\Psi$,
with $\hat{H}$ a quantum operator representing the reduced
Hamiltonian $\oint H_r$, if and only if $\Phi$ satisfies the
relation $i\d _t\Phi=\hat{H}^T\Phi$ where
\beq
\hat{H}^T=\hat{H}-\sqrt{2\pi}\e^{c_0}-i2\sqrt{2\pi} \d _{u' _0}
\eeq
is the quantum analogue of the Hamiltonian $H^T$
given by \eq (3.24). Therefore \eq (3.27) provides a well-defined
transformation between the physical states of the two considered
representations. Finally, it is not difficult to see that, for each
constant and real value of $t$, the states $\Psi$ and
$\Phi(c_0,u' _0=u_0+2\sqrt{2\pi}t,u_m,v_n)$ have the same
norm with respect to the inner product constructed in Ref.\,[18],
so that \eq (3.27) can in fact be interpreted as a unitary
transformation.

\subsection{The transform}

We will now consider the GWT for the Gowdy
model with the topology of a three-torus. After the
reduction (3.1)--(3.3) the infinitesimal generator $C$ of this
transform [given by \eq (2.16)] is expressed as
\beq
    C=\oint ({\cal K}E+\bar{K})  % 3.30
\eeq
where we have used that $k_{\theta\omega}=k_{\theta\nu}=0$ and
that $2\pi\omega,2\pi\nu\in S^1$. Taking then into account
\eq (3.25), the gauge fixing (3.13) and  (3.14) leads to
\beq
    C= -\oint H_r^{\prime(E)}=-H^T_{(E)}
\eeq
where the label $(E)$ refers to the Euclidean model. Notice in this
sense that, since the GWT acts in principle on functions
of the Euclidean phase space variables, mapping the Euclidean
constraints to the Lorentzian ones, the Hamiltonian density
appearing in \eq (3.31) must correspond to the Euclidean
theory [i.e., $\epsilon=1$ in \eq (3.26)]. Hence we conclude
that the infinitesimal generator of the GWT for the model
coincides, after gauge fixing, with minus the generator of the
dynamical evolution in Euclidean time $H^T_{(E)}$.

In the system reduction by the gauge-fixing procedure, Poisson
brackets transform into Dirac ones. As a consequence, for any
function $f$ that depends only on the reduced phase space
variables $(c_0,b_0' ,u' ,p_u,v,p_v)$ (and maybe on time), the
Poisson bracket $\{f,C\}$ becomes equal to $\{f,-H^T_{(E)}\}$.
Therefore, when acting on functions on the reduced phase space,
the action of the family of maps $W(t,t_0)$, introduced in \sect
2, turns out to be given by \eq (2.14) with $-H^T_{(E)}$
substituted for $C$. So far nothing has been specified about the
nature of the parameter $t$ of this family of maps. Our proposal
is to set this parameter equal to the time coordinate of our
gauge fixing. The motivation for this proposal comes from the
following considerations.

It is not difficult to check that, before imposing the gauge
conditions (3.13) and (3.14), the phase space of the Gowdy model
admits a canonical set of variables such that $z$ is the only
element that does not commute under Poisson brackets with the
infinitesimal generator of the GWT [given by \eq (3.30)]. For
instance, one can choose the variables $(u' =u+2z,\ p_u,\ v,\
p_v,\ w' =w-z,\
\bar{K},\ H' _r=-{\cal K}E-\bar{K},\ z)$ [see \eqs (3.6)--(3.9)
and (3.22)--(3.25)]. Taking then into account that $\e^z$ scales
as the densitized triad, we conclude that, before gauge fixing,
the effect of the GWT on any explicitly time-independent
function on phase space is a shift in the $z$-dependence that
replaces $z$ with $z+i\pi/2$. Once our time gauge is introduced,
the dependence on $z$ becomes an explicit dependence on the time
coordinate. Therefore, if the action of the GWT commutes with
the gauge-fixing procedure, the GWT for the gauge-fixed model
must be equivalent to a complex translation in the explicit time
dependence by a factor of $i\pi/2$ in which the implicit time
dependence is not modified. We will see that this is a result
obtained by identifying the time coordinate with the parameter
$t$ of the family of maps $W(t,t_0)$. Furthermore, we will show
in \sect 5 that in general relativity without symmetry
reductions or gauge fixing the action of the GWT amounts to that
of an inverse WR followed by a constant, complex conformal
transformation. With the proposed identification of parameters,
this is exactly the action that the GWT for our reduced model
will have on the Euclidean line element. Thus our proposal
guarantees that, on the line element, the process of model
reduction and gauge fixing commutes with the GWT. On the other
hand, we note that the proposed identification of parameters is
devoid of significance before gauge fixing, inasmuch as it leads
to the same formulae for the GWT as one obtains by assuming that
$t$ is just an abstract parameter on which the generator $C$
does not depend explicitly. With our gauge fixing, however, $C$
is replaced in \eq (2.14) by minus the reduced Euclidean
Hamiltonian that depends explicitly on time. Owing to this
dependence, \eq (2.15) [with $-H^T_{(E)}$ substituted for $C$]
is no longer valid for our reduced system. Employing instead \eq
(2.14), we arrive at the following action of the GWT on
functions on the reduced phase space:
\beq
W_{t_0}\circ f=\!\!\sum_{n=0}^{\infty}
\frac{(-1)^n}{n!}\,{\cal P}\left[\!\biggl\{f,\!\int_{t_0}^{t_0+i
\frac{\pi}{2}} H^T_{(E)}\biggr\}_{(n)}\!\right]\!
\eeq
where the path-ordered integration is over the explicit
$t$-dependence of $H^T_{(E)}$.

In particular, we see that the GWT for the reduced model depends
on the choice of the initial Euclidean time $t_0$. More
importantly, since $H^T_{(E)}$ is the generator of the dynamical
evolution, the map $W_{t_0}$ can be understood as an analytic
continuation in the complex plane of the time coordinate. In
fact, by defining a Lorentzian time through $t^{(L)}=t-i\pi/2$
(where $t$ is the Euclidean time), employing the relation
$H^T_{(E)}(t=t^{(L)}+i\pi/2)=H^T_{(L)}(t^{(L)})$ (which follows
from
\eq (3.26), with $H^T_{(L)}$ the reduced Lorentzian Hamiltonian), and
calling $\rho$ to the complete set of (implicitly time-dependent)
variables $(c_0,b' _0,u' , p_u, v, p_v)$, one can
check that the action of $W_{t_0}$ on any function $f(t_0,\rho)$ on
the reduced Euclidean phase space leads to the function on
the Lorentzian phase space obtained from $f$ by performing
the analytic continuation $t_0\to   t_0+i\pi/2$ just in the
explicit time dependence, while leaving unaltered the implicit
dependence on the time coordinate.

In quantum theory we expect the physical states $\Phi(t)$ of
the Euclidean $(E)$ and Lorentzian $(L)$ models to be related by the
transformation $\Phi^{(L)}(t_0)=\hat{W}_{t_0}\Phi^{(E)}(t_0)$.
According to \eq (2.20) and taking into account our identification
of parameters and the time dependence
of the Hamiltonian, the operator $\hat{W}_{t_0}$ should be
\beq  \hat{W}_{t_0}={\cal P}\left[\exp{\left(-i
\int_{t_0}^{t_0+i\frac{\pi}{2}}\hat{H}^T_{(E)}\right)}\right].
\eeq
It is not
difficult to prove [18] that the reduced Hamiltonian $H^T_{(E)}$
commutes under Poisson brackets with the only constraint of the
model $\Pi_0$, which is the same for the Euclidean and Lorentzian
systems. Let us then assume that in quantum theory $H^T_{(E)}$
is represented by an operator which commutes with the quantum
constraint $\hat{\Pi}_0$. Under this assumption, one can check that
the Lorentzian state $\Phi^{(L)}(t_0)=\hat{W}_{t_0}\Phi^{(E)}(t_0)$
is indeed annihilated by $\hat{\Pi}_0$ if $\Phi^{(E)}$ is in the
kernel of the constraint. Therefore the operator $\hat{W}_{t_0}$
provides a map between physical states.

Furthermore, if the Euclidean Hamiltonian $\hat{H}^T_{(E)}$ is
self-adjoint on the Hilbert space of physical states, we can
introduce a unitary evolution operator $\hat{U}_{(E)}$ in the
Euclidean quantum theory such that
$\Phi^{(E)}(t_1)=\hat{U}_{(E)}(t_1,t_0)\Phi^{(E)}(t_0)$. This
operator has the form
\beq
    \hat{U}_{(E)}(t_1,t_0)={\cal P}\left[\exp{\left(-i
            \int_{t_0}^{t_1}\hat{H}^T_{(E)}\right)}\right].
\eeq
Comparison of \eqs (3.33) and (3.34) allows one to interpret
$\hat{W}_{t_0}$ as the operator $\hat{U}_{(E)}(t_0+i\pi/2,t_0)$
corresponding to a continuation of the Euclidean time coordinate
from $t_0$ to $t_0+i\pi/2$.

Actually, the continuation $t_0\to t_0+i\pi/2$ transforms the
Euclidean model into the Lorentzian one. To see this, notice
that the two models would be completely equivalent if their
Hamiltonian densities $H_r' $ coincided. The only existing
constraint $\Pi_0$ is the same in both cases, and the Euclidean
and Lorentzian phase spaces are formally identical since they
both admit a set of elementary variables with the same
Poisson-bracket structure and domains of definition. On the
other hand, $H_r' $ given by \eq (3.26) is analytic in time,
while the constraint $\Pi_0$ and the elementary phase space
variables do not display any explicit time dependence. It is
then straightforward to check that the analytic continuation
$t_0\to t_0+i\pi/2$ made in the explicit time dependence maps
the reduced phase space, constraint and Hamiltonian of the
Euclidean theory to those of the Lorentzian case.

Substitution of $t_0$ by $t_0+i\pi/2$ in the explicit time
dependence turns out to relate as well the Euclidean and
Lorentzian line elements of the model. In Gowdy's time
coordinate $\tau=\e^t$, this substitution translates into
$\tau\to i\tau$, a transformation that resembles an
inverse WR [7]. Given \eq (3.21), the above substitution
sends the abstract Euclidean line element $ds^2_{(E)}$
to the Lorentzian one multiplied by a constant conformal factor,
namely $ds^2_{(E)}\to i ds^2_{(L)}$. Hence, from the
point of view of the set of four-geometries, the GWT can in fact be
interpreted as a composition of a complex rescaling of the metric
by a factor of $i$ and an inverse WR which maps the Euclidean
line element to its Lorentzian counterpart. This inverse WR can be
identified with the analytic continuation $\tau\to i\tau$
only if it is performed not in $ds^2_{(E)}$, but in
the explicit time dependence of the metric $ds^2_{(E)}/\tau$
which belongs to the same conformal equivalence class.

From the above discussion about the coincidence of the phase
spaces and constraints of the reduced Euclidean and Lorentzian
models, we expect that the corresponding quantum theories will
basically differ just in possessing distinct Hamiltonian and
evolution operators. The GWT, together with the Euclidean
unitary evolution, implies that the Lorentzian evolution
operator must be given by
\beq
\hat{U}_{(L)}(t_1,t_0)=\hat{W}_{t_1}\hat{U}_{(E)}
(t_1,t_0)\hat{W}^{-1}_{t_0}.
\eeq
Substituting \eqs (3.33) and (3.34), we then obtain that, at
least formally,
\beq
    \hat{U}_{(L)}(t_1,t_0)={\cal P}\left[\exp{\left(-i
        \int_{\Gamma}\hat{H}^T_{(E)}\right)}\right]
\eeq
where $\Gamma$ is the contour in the complex $t$-plane formed by
the three segments that join $t_0+i\pi/2$ with $t_0$, $t_0$ with
$t_1$, and $t_1$ with $t_1+i\pi/2$. Assuming that the Euclidean
Hamiltonian $\hat{H}^T_{(E)}$ depends analytically on $t$, as it
happens for its classical analogue, we can distort the contour
$\Gamma$ to the segment that goes from $t_0+i\pi/2$ to
$t_1+i\pi/2$. Defining then the Lorentzian Hamiltonian by
$\hat{H}^T_{(L)}(t^{(L)})=\hat{H}^T_{(E)}(t=t^{(L)}+i\pi/2)$,
we finally get
\beq
\hat{U}_{(L)}(t_1,t_0)={\cal P}\left[\exp{\left(
-i\int_{t_0}^{t_1}\hat{H}_{(L)}^T\right)}\right].
\eeq
Note that the definition adopted for $\hat{H}^T_{(L)}$ is just
a quantum version of the relation between the classical
Hamiltonians of the Euclidean and Lorentzian models.

In conclusion, we have shown that it is possible to find an
intrinsic time variable for the Gowdy model such that the action
of the GWT for the reduced system amounts to an analytic
continuation in the explicit time dependence. This continuation
maps the Euclidean line element to the Lorentzian one rescaled
by a factor of $i$. For other equivalent choices of intrinsic
time, of the form $t' =F(t)$, the GWT would clearly be
interpretable as the analytic continuation $t_0' =F(t_0)\to t_f'
=F(t_0+i\pi/2)$ performed in the explicit $t'$-dependence.
Although Gowdy's time (or any equivalent intrinsic time)
provides a natural and physically relevant choice of the time
coordinate for our model, there might also exist other
non-equivalent choices of time. One could then ask whether our
results would apply to these other possible choices of gauge as
well. It is not difficult to convince oneself that the purely
spacetime interpretation of the GWT as an analytic continuation
in the explicit time dependence will not be reachable in a
generic gauge. One can actually think of situations in which,
after a choice of intrinsic time which does not satisfy the
conditions that we have discussed in the Introduction, the
action of the GWT would imply an analytic continuation both in
the explicit and implicit time dependences. However, if the
action of the GWT on the abstract line element can be defined
before gauge fixing and the reduction of the system is
consistently performed in the gauge-fixing procedure, in the
sense that the GWT and the reduction commute (at least as far as
the line element is concerned), the conclusion that the effect
of the GWT on the Euclidean line element coincides with that of
an inverse WR and a complex conformal transformation should be
gauge-independent and therefore valid for the unreduced system.
We will see that this is indeed the case in \sect 5 where it
will be proved that for general relativity with no gauge fixing
the action of the GWT on the abstract line element is precisely
that mentioned above.

\section{Expanding universes}
\setcounter{equation}{0}

In order to show that the results obtained in the previous section
are not just an artifact of the particular gravitational system
studied, we will now discuss the relation between the GWT and the WR
for the case of expanding universes. By these, we will
understand spacetimes whose sections of constant time possess an
everywhere positive trace of the extrinsic curvature. Clearly, this
hypothesis about the expansion of the universe will not be generally
satisfied unless we restrict our attention to some specific class of
gravitational models. We will nevertheless carry out our analysis
without referring to any particular model reduction of general
relativity. It is worth remarking that the analysis of expanding
universes is of special interest for its applications in cosmology.
The feasibility and implications of our hypothesis of
expansion will be commented later in this section.

Let us start with the triad formulation of general relativity and
restrict our considerations to the sector of non-degenerate
three-metrics. We can first remove the gauge freedom associated with
the Gauss constraint by requiring that the densitized triad be
upper triangular. It is not difficult to check that this gauge
fixing is well-posed and consistent provided that the induced
metric is positive-definite. The Gauss constraint implies then
that the extrinsic curvature must be symmetric. The gravitational
system obtained with this gauge fixing can be described by the
variables
\bearr
        q^{ab}=\tilde{E}_i^a\tilde{E}^{bi}=h\,h^{ab},\nnn
    K_{ab}=K_a^i\; E_{_{_{\!\!\!\!\!\!\sim}}\;ib}=
            h^{-\frac{1}{2}}k_{ab},
\ear
for which the only non-vanishing Poisson brackets are
\beq
    \{q^{ab}(x),K_{cd}(y)\}=2\delta^{\;a}_{(c}
        \delta^{\;b}_{d)}\delta^{(3)}(x-y).
\eeq
Here the parentheses in the lower indices denote symmetrization.
Note that $q^{ab}$ and $K_{ab}$ are symmetric tensor densities of
weights 2 and $-1$, respectively. A canonical set of real
phase space variables is then
\bearr
    \frac{q^{11}}{2},\ K_{11},\ \frac{q^{22}}{2},\
            K_{22},\ \frac{q^{33}}{2},\ K_{33},q^{12}, \nnn
K_{12},\ q^{13},\ K_{13},\ q^{23},\ \ {\rm and\ }\  K_{23}.
\ear

The reality conditions $q^{ab}\in \R$ are not sufficient to
guarantee the positivity of $h^{ab}$; one must also require
\bearr
    \nq \det(q^{ab})\!=\!h^2\!>\!0,\ \
        q^{22}q^{33}-(q^{23})^2\!>\!0,\ \ q^{33}\!>\!0. \nnn
\ear
It is therefore most convenient to replace $q^{ab}$ by a new set
of configuration variables such that the sector of positive-definite
three-metrics corresponds to unrestricted real domains of
definition. This can be achieved with the change [21]
\bear
\nq u=\ln{q^{33}}-\frac{2}{3}\ln{h},\hspace*{1.85cm}& &
    \hspace*{.15cm}z=\frac{1}{3}\ln{h},\\
\nq v=\ln{\left[q^{22}q^{33}\!-\!(q^{23})^2\right]}\!-\!\frac{4}{3}
    \ln{h},&\;& \hspace*{.15cm}A=q^{12}h^{-\frac{2}{3}},\\
\nq B=q^{13}h^{-\frac{2}{3}},\hspace*{2.77cm}
    & & \hspace*{.15cm}C=q^{23}h^{-\frac{2}{3}}.
\ear
These equations can be viewed as part of a canonical
transformation in phase space. We will call the new canonically
conjugate momenta ($p_u,\ p_v,\ p_z,\ p_A$, $p_B$, $p_C$). An
explicit form of these momenta can be found in the Appendix
where we also display the inverse of our canonical
transformation. In particular, it turns out that
\beq
    p_z=K_{ab}q^{ab}=\sqrt{h} k_{ab}h^{ab},
\eeq
that is, $p_z$ is the densitized trace of the extrinsic curvature.

The restriction to the sector of positive-definite three-metrics is
now easily implemented by requiring that the new phase space
variables run over the whole real axis. On the other hand, defining
\bear
    \sigma \eql  p_u+2p_v+Ap_A+Bp_B+Cp_C,\\
U_a^{\,b}   \eql
    K_{ac}q^{cb}+\frac{\delta_a^b}{3}(2\sigma-p_z),\\
\bar{q}^{ab} \eql \e^{-2z}q^{ab},
\ear
we can write the vector constraint as
\beq
{\cal V}_a=\!-p_z\d _a z+\frac{2}{3}
\d _a(p_z+\sigma)\!-\!\d _b U_a^{\,b}+\bar{\Gamma}^c_{ab}U_c^{\,b}
\eeq
with $\bar{\Gamma}^c_{ab}$ being the Christoffel symbols of the
metric $\bar{q}_{ab}$ (the inverse of $\bar{q}^{ab}$).
The expressions for $U_a^{\,b}$ and $\bar{q}^{ab}$ in terms of our
new variables are given in the Appendix. A remarkable point is
that they are independent of the pair $(z,\ p_z)$. So all
the dependence of ${\cal V}_a$ on these two variables has been made
explicit in \eq (4.12).

It is straightforward to see that $\det \bar{q}^{ab} =1$.
As a particular consequence, we have $\bar{\Gamma}^b_{ba}=0$
(where summation over $b$ is understood).
Besides, one can check that $U_a^{\,b}$ is a tensor density of
weight 1. Therefore the last two terms in \eq (4.12) can in fact be
rewritten in the form $-\bar{D}_b U_a^{\,b}$, with $\bar{D}_a$ the
covariant derivative of the metric $\bar{q}^{ab}$.

In addition, the scalar constraint (2.7) can be written as
\beq
    {\cal S}=\frac{2}{3}\left(p_z^2-\Delta\right)
    -\frac{1}{\epsilon^2}\e^{3z}\;^{(3)}\!R,
\eeq
where
\beq
\Delta=\frac{1}{2}\left(3K_{ab}K_{cd}q^{ac}q^{bd} -p_z^2\right).
\eeq
An expression for $\Delta$ in terms of the variables (4.5)--(4.7)
and their momenta is obtained in the Appendix. It turns out that
$\Delta$ is a positive function on phase space which does not
depend on $z$ or $p_z$. On the other hand, we get from our
definitions that $h_{ab}=\e^z \bar{q}_{ab}$. Employing then the
familiar formula for the change of the curvature scalar under a
conformal transformation [16] and recalling that
$\bar{\Gamma}^b_{ba}=0$, we obtain
\beq
    \e^{z}\;^{(3)}\!R= ^{(3)}\!\bar{R}
        -2\d _b(\bar{q}^{ab}\d _a z)
            -\frac{1}{2}\bar{q}^{ab}\d _a z\d _b z,
\eeq
$^{(3)}\!\bar{R}$ being the curvature scalar of the metric
$\bar{q}_{ab}$. Substituting this relation into \eq (4.13), we arrive
at a formula for the scalar constraint in which all the dependence
on $z$ and $p_z$ is explicitly displayed.

Suppose now that we restrict our attention to a sector of
general relativity in which the trace of the extrinsic curvature
is everywhere nonzero. From \eq (4.8), this implies that the
momentum $p_z$ does not vanish at any point of the spacetime.
Using then the invariance of the theory under a time reversal,
we can limit $p_z$ to be positive without disregarding any
allowed geometry. In general, the classical solutions with
$p_z>0$ describe universes whose sections of constant time
expand with the evolution. An appealing choice of time gauge,
motivated by this expanding behaviour, is to set the metric
variable $z$ (which is an increasing function of the determinant
of the induced metric) equal to the time coordinate:
\beq
        z=t.
\eeq
Actually, it is easy to see that this condition fixes the gauge
freedom associated with the scalar constraint provided that
$p_z\neq 0$. Given the definition of $z$, our gauge fixing is
equivalent to setting $h^{1/3}=\tau=\e^t>0$. One can then easily
check that our choice of intrinsic time satisfies the conditions
which have been argued to guarantee that the GWT can be
interpreted as an analytic continuation in the explicit time
dependence after gauge fixing. On the other hand, since the
volume of the sections of constant time is proportional to
$h^{1/2}=\e^{3t/2}$, the choice of this volume as a
cosmological, intrinsic time for expanding universes is in fact
equivalent to our choice (4.16).

After gauge fixing, the dynamical evolution of the reduced
system is generated, modulo the vector constraint, by the
Hamiltonian $H^T=\int_{\Sigma}d^3x H_r$, where the reduced
Hamiltonian density $H_r$ is equal to minus the momentum $p_z$
obtained by solving the scalar constraint. From \eqs (4.13) and
(4.15), we get
\beq
    H_r=-p_z=-\sqrt{\Delta+\frac{3}{2\epsilon^2}
        \e^{2t}\,^{(3)}\!\bar{R}}.
\eeq
The square root in this formula is defined so that $\sqrt{1}=1$,
and we choose its cut to lie along the negative real axis.
On the other hand, with our gauge fixing, the vector constraint
(4.12) becomes
\beq  {\cal V}_a=\frac{2}{3}\d _a(\sigma-H_r)-
\bar{D}_bU_a^{\,b}.\eeq

Since $\Delta\geq 0$ on the phase space and $t\in \R$,
we see from \eq (4.17) that the requirement $p_z>0$ is
automatically satisfied in the Euclidean (Lorentzian) theory for
models in which $^{(3)}\!\bar{R}$ is strictly positive (negative).
Moreover, in this case the condition $p_z\neq 0$, which allows
the gauge fixing (4.16), is still verified under
an analytic continuation of the parameter $\epsilon$ as far as its
real (imaginary) part is kept positive (negative). In this sense,
for gravitational models with $^{(3)}\!\bar{R}>0$, the Lorentzian
Hamiltonian density $H_r^{(L)}$ could be defined as the limit of
the right-hand side of \eq (4.17) when $\epsilon$ tends to $-i$ from
the right half of the complex plane. This limit can be identified
with that of the Euclidean Hamiltonian density
$H_r^{(E)}\equiv H_r(\epsilon=1)$ when $t$ is replaced
by $t+i\eta$ and $\eta\to (\pi/2)^{-}$:
\beq
    H_r^{(L)}(t^{(L)})\equiv
    \lim_{\;\eta \to (\pi/2)^{-}}H_r^{(E)}(t= t^{(L)}+i\eta).
\eeq
Here $t^{(L)}$ denotes the Lorentzian time.
Similarly, for models with $^{(3)}\!\bar{R}<0$, we could define
\beq
    H_r^{(E)}(t)\equiv
    \lim_{\;\eta \to(\pi/2)^{-}}H_r^{(L)}(t^{(L)}=t-i\eta).
\eeq
Finally, the condition $p_z>0$ is also satisfied in models with
$^{(3)}\!\bar{R}=0$, both in the Euclidean and Lorentzian cases,
except if $p_u=p_v=p_A=p_B=p_C=0$, since it is only then that
$\Delta$ [given by \eq (A21)] vanishes. Our analysis would
therefore also apply to this class of gravitational systems if
the points with zero momenta could be suitably removed from the
reduced phase space.

For the expanding universes considered, we find from \eqs (2.16) and
(4.8) that, with our choice of gauge, the infinitesimal generator of
the GWT takes the form
\beq
    C=-\int_{\Sigma}d^3x\,H_r^{(E)}=-H^T_{(E)},
\eeq
$H_r^{(E)}$ being the Euclidean Hamiltonian density (4.17).
Thus, modulo the vector constraint, $C$ coincides with minus the
generator of the Euclidean dynamical evolution in the reduced
system. This relation is an analogue of \eq (3.31) for the Gowdy
model. Therefore, \eqs (3.32) and (3.33) are still valid for the
case of expanding universes if we follow again the proposal of
identifying our time coordinate with the parameter $t$ of the
family of maps (2.14). Moreover, using \eq (3.32) and applying a
line of reasoning similar to that discussed for the Gowdy model,
it is possible to show that, on functions $f$ on the reduced
phase space that are invariant under diffeomorphisms (so that
$\{f, {\cal V}_a\}=0$), the action of the GWT can be understood
as an analytic continuation in the explicit $t$-dependence. On
the other hand, assuming that our gauge fixing is well-posed in
the Euclidean sector and that the Hamiltonian $H^T_{(E)}$ can be
represented by a self-adjoint operator on the Hilbert space of
Euclidean physical states (that are annihilated by the vector
constraint), it is possible to introduce a unitary evolution
operator $\hat{U}_{(E)}$ in the Euclidean quantum theory through
\eq (3.34). When acting on Euclidean physical states, the
operator $\hat{W}_{t_0}$ of the GWT can then be interpreted as
the evolution operator $\hat{U}_{(E)}(t_0+i\pi/2,t_0)$ that
provides a continuation of the Euclidean time from $t_0$ to
$t_0+i\pi/2$.

In addition, let us suppose that the definition (4.19) is
meaningful, e.g., for models with $^{(3)}\!\bar{R}>0$ (or with
$^{(3)}\!\bar{R}=0$ if the points of the reduced phase space
with vanishing momenta can be consistently removed). Since the
only difference between the Lorentzian and Euclidean reduced
systems is found in the Hamiltonian density $H_r$ [see \eqs
(4.17) and (4.18)], we conclude that the classical Euclidean
theory can actually be mapped to the Lorentzian one by means of
the analytic continuation $t\to t+i\pi/2$, performed in the
explicit time dependence. This continuation must be understood
as a limit of the continuation $t\to t+i\eta$ when $\eta\in \R$
tends to $\pi/2$ from the left-hand side.

In quantum theory, the Lorentzian evolution operator would be
given again by \eq (3.35), or, equivalently, by (3.36). For models
with $^{(3)}\!\bar{R}>0$, e.g., the Euclidean Hamiltonian
$H^T_{(E)}$ is analytic with respect to its time dependence in the
whole region of the complex $t$-plane
\beq
\left\{t+i\eta\;:\;\,t\in \R,\  0\leq  \eta < \frac{\pi}{2}\right\}
\eeq
and possesses a well-defined limit when $\eta$ tends to $\pi/2$.
Provided that the quantum Hamiltonian $\hat{H}^T_{(E)}$ has also
this analytic behaviour, we can deform the contour of
integration $\Gamma$ in \eq (3.36) in such a way that it can be
taken as the limit of the segment that joins $t_0+i\eta$ with
$t_1+i\eta$ when $\eta\to (\pi/2)^{-}$, $t_0$ and $t_1$ being
the initial and final times appearing in the Lorentzian
evolution operator $\hat{U}_{(L)}$. In this way we arrive at
\beq
    \hat{U}_{(L)}(t_1,t_0)\!=\!\!\!\lim_{\,\eta
        \to\left(\frac{\pi}{2}\right)^{-}}\!{\cal P}\!
            \left[\exp{\!\left(-i\!\int_{t_0+i\eta}^{t_1+i\eta}
                \!\!\!\hat{H}^T_{(E)}\!\right)}\right]\!,
\eeq
which coincides formally with the operator
\beq
     {\cal P}\left[\exp{\left(-i\int_{t_0}^{t_1}
    \hat{H}^T_{(L)}\right)}\right]
\eeq
if we define the quantum Lorentzian Hamiltonian $\hat{H}^T_{(L)}$
by analogy with \eq (4.19):
\beq  \hat{H}^T_{(L)}(t^{(L)})=\!\lim_{\,\eta
\to\left(\pi/2\right)^{-}} \hat{H}^T_{(E)}
(t=t^{(L)}+i\eta).\eeq

Let us finally analyze the transformation of the line element under
the substitution of the Euclidean time $t$ by $t+i\pi/2$ in the
explicit time dependence. We have
seen that, before fixing the time gauge, the induced metric is
given by $h_{ab}=\e^z\bar{q}_{ab}$, $\bar{q}_{ab}$ being the inverse
of the metric (A1)-(A2), which is independent of $z$ and $p_z$.
On the other hand, the compatibility of our gauge-fixing condition
with the dynamical evolution requires that
\bear
    1=\frac{dz}{dt} \eql \left\{z,\int_{\Sigma}d^3x\left(
    \frac{N_{_{_{\!\!\!\!\!\!\sim}}\;}}{2}{\cal S}-N^a{\cal V}_a
                        \right)\right\} \nn
    &\approx&-\frac{2}{3}N_{_{_{\!\!\!\!\!\!\sim}}\;}H_r+\frac{2}{3}
        \d _a N^a,
\ear
where ${\cal V}_a$ and ${\cal S}$ are the constraints (4.12) and
(4.13), the symbol $\approx$ denotes equality modulo these
constraints and \eq (4.16), and $H_r$ is given by \eq (4.17).
Since the lapse function is
$\sqrt{h}N_{_{_{\!\!\!\!\!\!\sim}}\;}$, with $h=\e^{3z}$, we find
that, for $z=t$, the line element can be written as
\bearr
    ds^2=\frac{\e^{3t}}{[\epsilon\, H_r]^2}
        \,\left(\frac{3}{2}-\d _a N^a\right)^2 dt^2 \nnn
\cm     +\;\e^t\;\bar{q}_{ab}\;(dx^a+N^adt)\;(dx^b+N^bdt).
\ear
Here we have assumed that our gauge fixing is
acceptable, at least in the sense of a limit (as discussed above for
gravitational models with definite sign of the curvature scalar
$^{(3)}\!\bar{R}$), both in the Euclidean and Lorentzian cases.

From \eqs (4.17) and (4.27) we conclude that the substitution of
$t$ by $t+i\pi/2$ (in the explicit time dependence) sends the
Euclidean line element to the Lorentzian one multiplied by a
factor of $i$, as it happened in the Gowdy model. Therefore we
again arrive at the conclusion that the action of the GWT on the
set of Euclidean four-geometries can be interpreted as a
composite of a conformal transformation by a constant complex
factor and an inverse WR. This WR can be performed by
substituting the Euclidean time $\tau=\e^t$ with $i\tau$ in the
explicit time dependence of the metric $ds^2_{(E)}/\tau$, which
is a representative of the conformal equivalence class to which
the Euclidean metric $ds^2_{(E)}$ belongs. As to the validity of
our results for other possible choices of gauge, comments
similar to those presented at the end of \sect 3 apply to the
present case as well.

\section{The transform in vacuum general relativity}
\setcounter{equation}{0}

We will now study the action of the GWT and its relation with
the WR in vacuum general relativity without imposing any gauge
fixing nor symmetry or model reduction. In this general case, we
do not have at our disposal any natural choice of time as a
function of the gravitational phase space variables. This
prevents us from identifying the infinitesimal generator of the
GWT with the generator of the dynamical evolution in a
particular time variable. Therefore one cannot achieve for the
GWT the purely spacetime interpretation of just an analytic
continuation in the explicit dependence on a suitable time
coordinate. However, we will see that it is still possible to
relate the GWT with an inverse WR. As we commented in the
Introduction, it was initially far from clear that a relation
between these two transformations could actually exist [5,6].
Our purpose in this section is to clarify this point and to show
that the action of the GWT can in fact be interpreted as the
result of performing an inverse WR and a constant, complex
conformal transformation. Recall that we regard the WR for
vacuum general relativity as a series of transformation rules
for the lapse, shift and gravitational phase space variables
that map the Lorentzian to the Euclidean abstract line element,
and the Lorentzian action to the Euclidean one rescaled by a
factor of $i$.

Let us start our discussion by analyzing the
behaviour of the line element under the transform introduced by
Thiemann. From \eqs (2.2) and (2.17) we obtain
\beq
W\circ h^{ab}=W\circ \left(\frac{\tilde{E}^a_i
\tilde{E}^{bi}}{{\rm det}(\tilde{E}^c_j)}\right)=-i h^{ab}.
\eeq
So the action of the GWT on the induced metric is simply a
constant, complex rescaling: $W\circ h_{ab}=i h_{ab}$. On the other
hand, we recall that the GWT is in principle defined as a map on
functions on phase space, specially designed to send the Euclidean
constraints (2.3) and (2.7) (with $\epsilon=1$) to their Lorentzian
counterparts. For functions which do not depend on the phase space
variables, it is nevertheless trivial to extend the definition of
the GWT by using \eq (2.14). On this kind of functions, the GWT
acts then as the identity operator. It therefore seems natural
to assume that the densitized lapse function and the shift vector
are invariant under the GWT,
\beq
    W\circ N_{_{_{\!\!\!\!\!\!\sim}}\;}=
    N_{_{_{\!\!\!\!\!\!\sim}}\;},\;\;\;\;\;\;\;\;W\circ N^a=N^a
\eeq
because $N_{_{_{\!\!\!\!\!\!\sim}}\;}$ and $N^a$ are not phase space
variables, but just the Lagrange multipliers of the scalar and
vector constraints. A different way to arrive at these
transformation rules is to demand that the action of the GWT is
consistent with the expression of the extrinsic curvature [16]
\beq
    k_a^{\,b}=\frac{h^{bc}}{2N}\left[\d _t h_{ac}-D_a(N^d h_{cd})-
    D_c(N^d h_{ad})\right].
\eeq
Recalling that the lapse function is given by
$N=\sqrt{h}N_{_{_{\!\!\!\!\!\!\sim}}\;}$ and using \eq (5.1), we
conclude that the GWT sends the right-hand side of \eq (5.3) to
\bearr
    \frac{1}{i\sqrt{i}} \frac{h^{bc}}{2\sqrt{h}
    \{W\circ N_{_{_{\!\!\!\!\!\!\sim}}\;}\}}\left[\;\d_t h_{ac}
        -D_a(\{W\circ N^d\}h_{cd})\right. \nnn \inch
    -\left.D_c(\{W\circ N^d\}h_{ad})\;\right].
\ear
In addition, \eqs (2.2) and (2.17) imply
\beq
     W\circ k_a^{\,b}=W\circ (h^{-\frac{1}{2}} K_a^i
        \tilde{E}^a_i)=\frac{1}{i\sqrt{i}}\,k_a^{\,b}.
\eeq
One can then readily see that the expression (5.4) reproduces the GWT
of the extrinsic curvature for any choice of densitized lapse
function and shift vector if and only if \eq (5.2) is satisfied.

Except for a sign in the transform of the densitized lapse, \eq
(5.2) reproduces the transformation rules deduced in Ref.\,[6]
by imposing that the GWT maps the Euclidean action
to the Lorentzian one. This difference of a sign is due to
the unusual convention adopted in Ref.\,[6] for the sign of the
Euclidean gravitational action, which is the opposite of the
standard one. There has also been some confusion [6] about a
possible interpretation of the transformation rules (5.2).
We will show at the end of this section that such rules
are precisely those that one would obtain with the continuation
$N\to iN$ (which corresponds to an inverse WR)
when this is composed with the
conformal transformation $ds^2\to ids^2$.

On the other hand, making use of \eqs (5.1) and (5.2), we get
\beq
    W\circ N= i \sqrt{i} N.
\eeq
It is now straightforward to check that the abstract Euclidean
line element,
\beq
    ds^2_{(E)}\!=\!N^2dt^2\!+h_{ab}(dx^a+N^adt)(dx^b+N^bdt),
\eeq
is mapped by the GWT to
\beq
    W\circ ds^2_{(E)}=ids^2_{(L)}
\eeq
where $ds^2_{(L)}$ is the Lorentzian metric. Therefore, from the
point of view of the set of four-geometries, the GWT can be
considered as equivalent to an inverse WR that
sends the Euclidean line element to the Lorentzian one, followed by
a conformal transformation of the metric,
$ds^2_{(L)}\to \Omega^2 ds^2_{(L)}$, in which the
squared conformal factor is constant and imaginary, namely,
$\Omega^2=i$.

The geometric interpretation that has been attained for the GWT
can also be achieved by analyzing the relation between the
Euclidean and Lorentzian actions of the geometrodynamic
formulation. From \eq (5.1) and the formula for the curvature
scalar of a conformally transformed metric [16] it follows that
$W\circ\,^{(3)}\!R(h_{ab})=-i\,^{(3)}\!R(h_{ab})$. Employing \eq
(5.5), it is then a simple exercise to check that the GWT maps
the Euclidean action (1.2) to its Lorentzian counterpart (1.1):
\beq
    W\circ I_{(E)}= S_{(L)}.
\eeq

In agreement with our discussion above, this
correspondence between the geometrodynamic actions can be
interpreted as that obtained by composing an inverse WR
with a conformal transformation of the Lorentzian metric when
$\Omega^2=i$. A conformal transformation with the factor $\Omega$ has
the following action on the metric functions:
\beq
    h_{ab}\to \Omega^2 h_{ab},\;\;\;\;\;
    N\to \Omega N,\;\;\;\;\;N^a\to N^a.
\eeq
The corresponding change in the extrinsic curvature can be
computed from \eq (5.3). Provided that $\Omega$ is constant,
since the Christoffel symbols are invariant under a constant
rescaling of the metric, we arrive at
\beq
    k_a^{\,b}\to \Omega^{-1} k_a^{\,b}.
\eeq
Finally, for constant $\Omega$, the change in the curvature
scalar of the induced metric is the familiar one: $^{(3)}\!R\to
\Omega^{-2}\, ^{(3)}\!R$. One can then see that a constant
conformal transformation results in multiplying the Lorentzian
action (1.1) by a factor of $\Omega^2$. Since an inverse WR maps
the Euclidean action $I_{(E)}$ to $-iS_{(L)}$, the composite of
this rotation and a constant conformal transformation turns out
to send the action $I_{(E)}$ to $-i\Omega^2 S_{(L)}$. Thus, for
$\Omega^2=i$, we recover the transformation rule (5.9).

One may wonder whether the relation found between the
GWT and the inverse WR in vacuum gravity continues to be valid
when one compares the effects of these two transformations not
just on the abstract line element and the gravitational action, but
on any function of the shift vector, densitized lapse function, and
gravitational phase space variables. Actually, the answer to this
question turns out to be in the affirmative provided that the
inverse WR is implemented by making the analytic continuation
$N\to iN$.

We will prove this statement in the triad formulation of vacuum
general relativity, although our discussion can be easily extended
to the Ashtekar and geometrodynamic formulations. Let us first
consider the action of an inverse WR obtained by performing
an analytic continuation of the (Euclidean) lapse function from
positive real to positive imaginary values. Denoting this action
by $R^{-1}$, we then have $R^{-1}\circ N=iN$
(with $N\in \R^+$). Since an inverse WR maps
Euclidean metrics to Lorentzian ones, we also get
\beq
    R^{-1}\circ N^a=N^a,\;\;\;\;\;
    R^{-1}\circ \tilde{E}^a_i=\tilde{E}^a_i.
\eeq
In particular, this equation implies
$R^{-1}\circ h_{ab}=h_{ab}$. Using now the definition of the
densitized lapse function and \eqs (2.2) and (5.3), one arrives at
\beq
    R^{-1}\circ N_{_{_{\!\!\!\!\!\!\sim}}\;}= i
    N_{_{_{\!\!\!\!\!\!\sim}}\;},\;\;\;\;\;\;\;
    R^{-1}\circ K_a^i=-iK_a^i.
\eeq
On the other hand, the action of a conformal transformation on
the metric functions and the extrinsic curvature is given by
\eqs (5.10) and (5.11). For $\Omega^2=i$, the corresponding constant
rescaling of the shift vector, densitized lapse function, and
phase space variables of the triad formulation is
\beq
    N^a\to N^a,\cm  N_{_{_{
    \!\!\!\!\!\!\sim}}\;}\to -iN_{_{_{\!\!\!\!\!\!\sim}}\;},
\eeq
\beq
    \tilde{E}^a_i\to i\tilde{E}^a_i,\cm     K_a^i\to K_a^i.
\eeq
It then follows that the composite of an inverse WR (obtained
with the prescription $N\to i N$) and a conformal transformation
with factor $\Omega^2=i$ preserv\-es the Poisson brackets of
vacuum gravity and leaves $N^a$ and
$N_{_{_{\!\!\!\!\!\!\sim}}\;}$ invariant. Moreover, this
composite transformation has in fact the same effect as the GWT
on the gravitational variables, shift, and densitized lapse, as
can be seen by comparing \eqs (5.12)--(5.15) with \eqs (2.17)
and (5.2). We hence conclude that the action of the GWT on
functions on phase space that depend on the shift vector and the
densitized lapse function can indeed be interpreted in vacuum
general relativity as the result of an inverse WR and a
constant, complex conformal transformation.

\section{Conclusions and further comments}
\setcounter{equation}{0}

We have analyzed whether there exists a relation between the GWT
constructed by Thiemann and the WR employed in the
geometrodynamic (or triad) formulation of vacuum gravity, an
issue which remained obscure in the literature [5,6]. It has
been seen that the answer is in the affirmative. More precisely,
the transformation rules for the lapse, shift and gravitational
phase space variables are the same under the GWT as under an
inverse WR composed with a conformal transformation of a
(squared) factor equal to $i$, provided that the WR is
implemented by means of a complex rotation of the lapse
function. We have also discussed whether, in special
circumstances, the GWT admits the simple spacetime
interpretation of an analytic continuation in time. It has been
proved that, in certain systems which possess an intrinsic time
variable which behaves as a logarithm of the densitized triad
under constant scale transformations, the GWT reduces, after
fixing the time gauge, to an analytic continuation that is
performed only in the explicit dependence on the time
coordinate.

Let us describe in more detail the results obtained. We have
first considered two different types of gravitational models,
namely, the Gowdy model with the topology of a three-torus and a
class of gravitational systems which represent universes with
expanding sections of constant time. In these models, the
dynamical evolution allows one to choose, as a time coordinate,
a function of the induced metric which is not invariant under
constant scale transformations. After a partial gauge fixing
which incorporates this particular choice of time, the
infinitesimal generator of the GWT turns out to coincide (modulo
first-class constraints) with minus the reduced Hamiltonian of
the system. Given this result, we have proposed to identify the
parameter $t$ employed to define the GWT with the introduced
time coordinate. This proposal has been seen to guarantee that
(at least on the abstract line element and functions on phase
space which are explicitly time-independent before fixing the
time gauge) the action of the GWT and the gauge-fixing procedure
commute. Furthermore, it then follows that, when acting on
functions on the partially reduced phase space that commute with
the existing first-class constraints, the GWT for the reduced
system can be interpreted as an analytic continuation made only
in the explicit time dependence. In addition, the operator that
implements this GWT in the quantum theory is then formally
identical to the evolution operator which, on physical states,
would result in a continuation of the Euclidean time $t$ until
it acquires a constant, imaginary part equal to $\pi/2$. For the
models analyzed, we have also shown that this continuation in
the explicit time dependence maps in fact the Euclidean theory
to the Lorentzian one, because it sends the reduced phase space,
first-class constraints and the reduced Hamiltonian of the
Euclidean systems to their Lorentzian counterparts.

In these models, the substitution of $t$ by $t+i\pi/2$ in the
explicit time dependence has been seen to transform the
Euclidean line element into the Lorentzian one, though
multiplied by a factor of $i$. This constant factor is
irrelevant in what concerns the Einstein equations. However, its
appearance in the Lorentzian line element reveals that the GWT
implies a complex conformal transformation of the metric. On the
other hand, we have argued that the transformation rule of the
Euclidean line element under the GWT, namely $ds^2_{(E)}\to i
ds^2_{(L)}$, must in fact be gauge-independent. In this way, we
have arrived at a geometric interpretation of the GWT as the
composite of an inverse WR (which maps the set of Euclidean
four-geometries to its Lorentzian counterpart) and a complex
conformal transformation with $\Omega^2=i$.

In general relativity one cannot introduce a well-posed and
globally defined time gauge. Hence, one loses a purely spacetime
picture for the GWT as an analytic continuation in the explicit
dependence on the time coordinate. Nevertheless, we know that
this transform sends the Euclidean first-class constraints to
those of the Lorentzian sector. In addition, we have seen that,
in vacuum general relativity with neither gauge fixing nor
symmetry reduction, the transformation rules for the Euclidean
action and the abstract line element under the GWT are formally
the same as one obtains by composing an inverse WR and a
constant conformal transformation, provided that $\Omega^2=i$.
Furthermore, a similar result has been obtained when discussing
the transformation rules for the gravitational phase space
variables, the shift vector, and the densitized lapse function,
once we have adhered to the prescription of implementing the WR
by means of an analytic continuation of the lapse function.
Therefore, we conclude that, on functions on phase space that
depend as well on the shift and the densitized lapse, the action
of the GWT for vacuum gravity can actually be interpreted as the
result of performing an inverse WR and a suitable conformal
transformation.

The fact that the GWT can be related to the WR in vacuum gravity
allows one to regard the transform introduced by Thiemann as a
rigurous procedure to carry out the WR in general relativity, a
rotation that would otherwise remain a rather formal technique
for passing from the Lorentzian gravitational theory to the
Euclidean one. For all possible applications of the GWT, one
must nonetheless keep in mind that the action of this transform
cannot be identified just with that of a pure (inverse) WR
because it involves in addition a complex rotation of the
conformal factor.

At first sight, it might seem striking that the GWT relates the
Euclidean and Lorentzian theories while mapping the Euclidean
action $I_{(E)}$ to the Lorentzian one, $S_{(L)}$, since in the
path-integral approach to geo\-me\-tro\-dynamics one usually
assumes that the Lorentz\-ian weight $\e^{iS_{(L)}}$ must be
replaced by $\e^{-I_{(E)}}$ when dealing with Euclidean
histories. However, if one really considers pure Euclidean
gravity rather than a Euclidean version of Lorentzian general
relativity, the weight for a Euclidean geometry in the path
integral should indeed be given by $\e^{iI_{(E)}}$. In this
sense, a correct substitution should be the one supplied by the
GWT, i.e., $I_{(E)}\to S_{(L)}$. On the other hand, if it were
possible to arrive at a convergent Euclidean path integral for
gravity, the conformal factor $\Omega$ ought to be integrated
over a complex contour [7,11]. But once the conformal factor is
taken as a complex variable, the sums over conformally
transformed Euclidean metrics obtained with the weights
$\e^{iI_{(E)}}$ and $\e^{-I_{(E)}}$ become equivalent because
the Euclidean action $I_{(E)}$ gets multiplied by a factor of
$i$ under the conformal transformation $ds^2_{(E)}\to i
ds^2_{(E)}$. So, the two considered weights differ only by a
constant rotation of $\pi/4$ in the complex $\Omega$-plane. We
thus see that, modulo complex rotations of the conformal factor,
the GWT and the WR lead to equivalent prescriptions for the path
integral over Euclidean geometries. The use of the GWT has
nonetheless clear advantages since it is formulated on a much
more rigurous basis than the WR for gravity, and its application
in order to extract the Lorentzian physics from Euclidean
general relativity is in principle independent of the
quantization approach followed and does not require the
availability of a classical notion of spacetime.

In this paper we have restricted our analysis of the GWT to the
case of vacuum general relativity. Some of the results obtained here
might be modified by the existence of a matter content. The physical
implications that the existence of a Wick transform may have for
gravity with matter fields and a cosmological constant [6] will be
the subject of future research.

\section*{Appendix}
\setcounter{equation}{0}
\renewcommand{\theequation}{A\arabic{equation}}

In this appendix we explicitly construct the canonical
transformation introduced in \sect 4, and obtain the expressions
for $\bar{q}^{ab}$, $U_a^{\,b}$ and $\Delta$ in terms of the new
phase space variables.

The inverse of \eqs (4.5)--(4.7) is given by $q^{ab}=\e^{2z}
\bar{q}^{ab}$,
where
\bearr
\bar{q}^{11}=(1+B^2\e^{-u+v}+Y^2)\e^{-v},
\hspace*{.4cm}\;\;\;\;\;\bar{q}^{22}=X,\yyy
\bar{q}^{33}=\e^u,\;\;\;\;\;\bar{q}^{12}=A,\;\;\;\;\;
\bar{q}^{13}=B,
\;\;\;\;\;\bar{q}^{23}=C.
\ear
Here we have adopted the notations
\beq
    X=(\e^v+C^2)\e^{-u},\qquad    Y=A\e^{\frac{u}{2}}-
        BC\e^{-\frac{u}{2}}.
\eeq
In particular, we see that $\bar{q}^{ab}$ is independent of $z$.

The above relations are part of a canonical transformation generated
by the functional
\beq
    {\cal F}=-\frac{1}{2}\int_{\Sigma}K_{ab}
    q^{ab}(u,v,z,A,B,C)
\eeq
where $q^{ab}(u,v,z,A,B,C)$ denote the functions (A1)-(A2) of the
new configuration variables. The new momenta can then be
straightforwardly computed from the generating functional
${\cal F}$. Using \eqs (A1)-(A4), we get
\bearr
    p_u=\frac{\e^{2z}}{2}(A^2\e^u-XB^2)\e^{-v}K_{11}
\nnn
    \inch  +\;\frac{\e^{2z}}{2}\left[-XK_{22}+
                    \e^uK_{33}\right],\yyy
p_v=\frac{\e^{2z}}{2}\left[-(1+Y^2)
        \e^{-v}K_{11}+\e^{-u+v}K_{22}\right],\yyy
p_A=\e^{2z}
     \left[Y\e^{\frac{u}{2}-v}K_{11}+K_{12}\right],\yyy
p_B=
       \e^{2z}\left[(XB-AC)\e^{-v}K_{11}+K_{13}\right],\yyy
p_C=\e^{2z}\left[-YB\e^{-\frac{u}{2}-v}
                    K_{11}+C\e^{-u}K_{22}+K_{23}\right]
\ear
and
\beq
    p_z=K_{ab} q^{ab}(u,v,z,A,B,C).
\eeq

After some calculations one finds that the inverse of these
formulae is
\bearr
        K_{11}=\e^{-2z+v}\,\Theta,\yyy
K_{22}=\e^{-2z+u-v}\left[(1+Y^2)\,\Theta+2p_v
            \right],\yyy
K_{33}=\e^{-2z}\left[(1+Y^2+B^2\e^{-u+v})X\e^{-v}-
            A^2\right]\Theta       \nnn
        \cm \cm  +\e^{-2z}\left[2\e^{-u}p_u+2X\e^{-v}p_v\right],
                    \yyy
K_{12}=\e^{-2z}\left[-Y\e^{\frac{u}{2}}\Theta+p_A
                \right],\yyy
K_{13}=\e^{-2z}\left[(AC-XB)\,\Theta+p_B\right],
                 \yyy
K_{23}=\e^{-2z}\left[YB\e^{-\frac{u}{2}}-(1+Y^2)C
            \e^{-v}\right]\Theta\nnn
         \cm\cm + \e^{-2z}\left[p_C-2C\e^{-v}p_v\right],
\ear
where
\beq
    \Theta=\frac{1}{3}(p_z-2\sigma)
\eeq
and $\sigma$ is the function defined in \eq (4.9).

Substituting \eqs (A11)--(A16) into the right-hand side of
\eq (4.10), it is possible to show that
\bearr
    U_1^{\,b}=\bar{q}^{2b}p_A+\bar{q}^{3b}p_B,\yyy
    U_2^{\,b}=\bar{q}^{1b}p_A\!+\bar{q}^{3b}p_C\!+
        2(\bar{q}^{2b}\e^u-\bar{q}^{3b}C)\e^{-v}p_v, \yyy
U_3^{\,b}=
    \bar{q}^{1b}p_B+\bar{q}^{2b}p_C+2\bar{q}^{3b}\e^{-u}p_u\nnn
\cm\cm  +\;2\;(\bar{q}^{3b}X-\bar{q}^{2b}C)\;\e^{-v}p_v.
\ear
All elements of the matrix $U_a^{\,b}$ are therefore independent
of the canonical pair of variables $(z,p_z)$.

Finally, an expression for $\Delta$ in terms of our new phase space
variables can be obtained from \eqs (4.14), (A1)--(A3) and
(A11)--(A17) after a lengthy computation. The result can be written
in the form
\bearr
\Delta= \sigma^2\!+3\left[p_u\!+Bp_B\!+Cp_C\!+
(2BC\e^{-u}\!-\!A)p_A\right]^2
    \nnn
\ \ + 3 \e^v\left[p_C+Y\e^{\frac{u}{2}-v}p_B+(YC\e^{\frac{u}{2}-v}
+B)\e^{-u}p_A\right]^2
    \nnn
\ \ + 3\e^{u-v}\left[p_B+C\e^{-u}p_A\right]^2+
            3\e^{-u}p_A^2.
\ear
Thus $\Delta$ is a positive function on phase space that
depends on neither $z$ nor $p_z$.

\acknowledgments

The author is very grateful to P. F. Gonz\'{a}lez D\'{\i}az for valuable
comments and discussions. He is also thankful to A. Ashtekar for
helpful conversations.


\begin{thebibliography}{99}
\bibitem{17}
All integrals in this equation are well defined if the function $C$
depends analytically on the integration parameter in a complex
region which contains the segment joining $t_0$ and $t$.
\end{thebibliography}
\end{document}